\documentclass[aip,amsmath,amssymb,reprint,]{revtex4-1}

\usepackage{graphicx}% Include figure files
\usepackage{dcolumn}% Align table columns on decimal point
\usepackage{bm}% bold math
\usepackage[utf8]{inputenc}
\usepackage[T1]{fontenc}
\usepackage{mathptmx}
\usepackage{etoolbox}
\usepackage{float}
\usepackage{placeins}

\makeatletter
\def\@email#1#2{%
 \endgroup
 \patchcmd{\titleblock@produce}
  {\frontmatter@RRAPformat}
  {\frontmatter@RRAPformat{\produce@RRAP{*#1\href{mailto:#2}{#2}}}\frontmatter@RRAPformat}
  {}{}
}%
\makeatother
\begin{document}

\preprint{AIP/123-QED}

\title[Critical parameters of an oval billiard with an elliptical component]{Critical parameters of an oval billiard with an elliptical component}

\author{Anne Kétri P. da Fonseca}
\affiliation{Departamento de Física, UNESP - Universidade Estadual Paulista, Rio Claro, SP, Brazil}
\email{anne.ketri@unesp.br}

\author{Joelson D. V. Hermes}
\affiliation{Federal Institute of Education, Science and Technology of South of Minas Gerais - IFSULDEMINAS, Inconfidentes, MG, Brazil}
\affiliation{Physics Institute, University of São Paulo - USP, 05508-090, São Paulo, SP, Brazil.}

\author{Edson D.\ Leonel}
\affiliation{Departamento de Física, UNESP - Universidade Estadual Paulista, Rio Claro, SP, Brazil}

\date{\today}

\begin{abstract}
We explore the critical parameters responsible for the transition from integrability to chaos in a family of billiards combining elliptical and oval deformations. Unlike standard oval billiards, where a known critical parameter governs the destruction of the last invariant curve, the introduction of an integrable elliptic component yields a second deformation axis. We derive an analytical expression for the critical parameter $\varepsilon_c$ in this combined system and validate it numerically using Slater’s theorem, showing that increasing the elliptical component lowers the critical threshold for global chaos. Moreover, we uncover a previously unexplored regime: when the two deformation components are in phase, the elliptic contribution progressively suppresses chaos, leading to the restoration of invariant curves and periodic orbits. A first-order analytical approximation confirms this behavior, supported by numerical validation. Our results reveal how the interplay between distinct boundary deformations enriches phase-space organization and offers enhanced controllability of chaotic dynamics in billiard systems.
\end{abstract}

\maketitle

\begin{quotation}
The transition from integrability to chaos in dynamical systems is typically governed by critical parameters. In billiards, these are often tied to boundary deformations. For the oval billiard, an analytical expression is known for the critical parameter where convex pieces are introduced in the boundary, leading to the destruction of the last invariant curve an the onset of global chaos. The ovoid elliptic billiard introduces a second deformation component given by the integrable elliptical component. This interplay gives rise to novel dynamics on the phase space. We obtain an analytical expression for the critical parameter $\varepsilon_c$ in this combined system and validate it numerically through method based on Slater's theorem. Both results indicate the decrease of the critical value where the last invariant curve is destroyed as the elliptical component is increased. Furthermore,when in phase, the elliptic deformation progressively suppresses the chaotic sea introduced by the oval component, restoring invariant curves and periodic orbits. A first-order approximation is obtained, confirming this scenario and also validated by the same numerical method. These results provide deeper insight into how boundary deformations shape phase-space structures in billiards, while offering enhanced tunability through multiple control parameters.  
\end{quotation}

\section{\label{Introduction} Introduction}
Billiard systems play an important role as idealized systems of non-interacting articles confined to a bounded region \cite{chernov}. The shape of the boundary of this confinement determines the dynamical properties of the system, ranging from fully integrable to chaotic or mixed behavior. Moreover, the many possibilities of describing and deforming these boundaries enables numerous practical applications in areas such as celestial mechanics \cite{celestial}, plasma physics \cite{plasma}, photonics \cite{17,26}, and condensed matter physics \cite{29}. Time dependence and dissipation can also be incorporated into these systems, further expanding the range of applications and allowing for a more adequate representation of observed physical phenomena \cite{6,livrodenisspringer}.

\begin{figure}[h]
\includegraphics[width=\columnwidth]{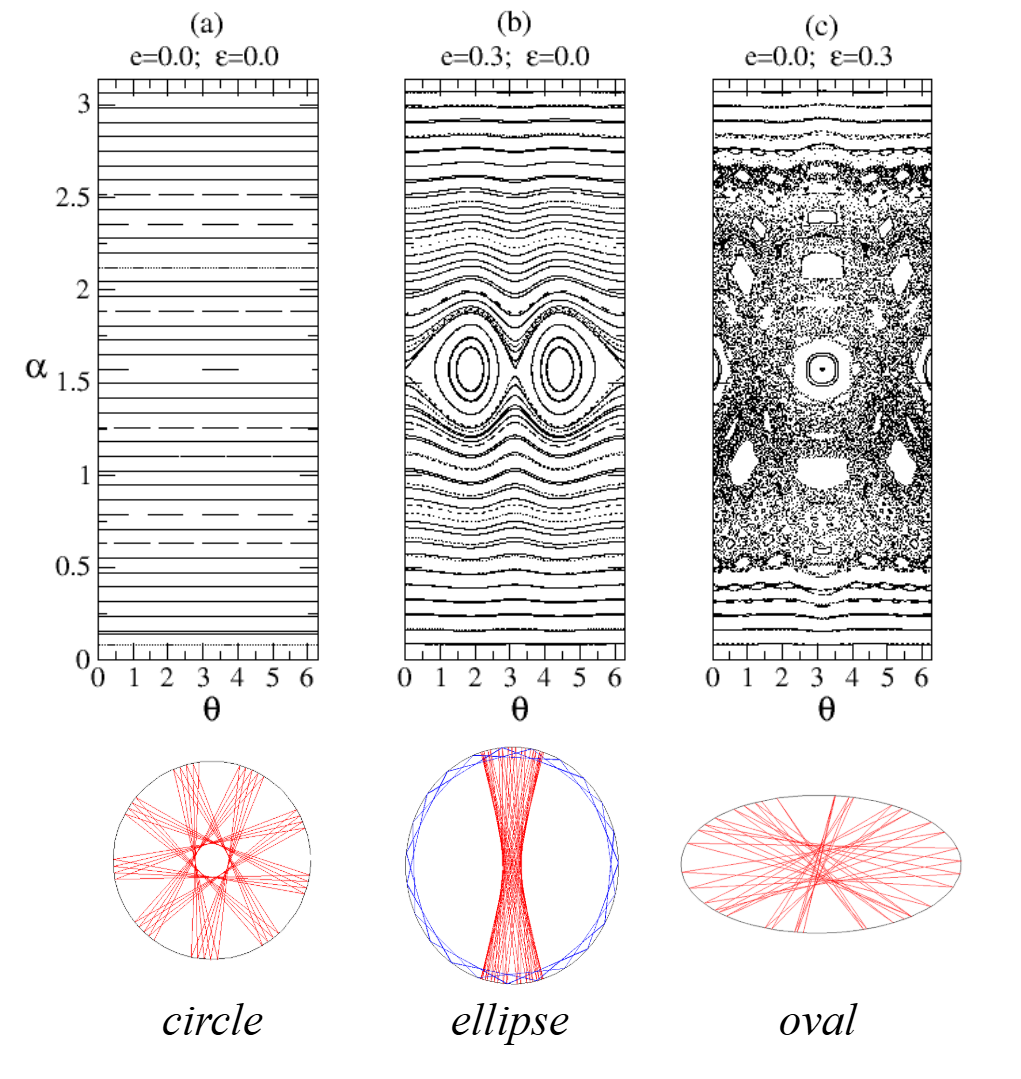}% Here is how to import EPS art
\caption{\label{Fig1} Phase space, boundary and orbit for a single particle for the (a) circular billiard; (b) elliptical billiard ($q=1$) with the rotation orbit in blue and the libration orbit in red and (c) oval billiard for $p=1$. Control parameters associated with Eq.~(1) are indicated in the figure.}
\end{figure}

Formally, these systems fall within the Hamiltonian  $H(x,p,t)={\mathcal{P}}^2/2m+ V(x,t)$. We restrict our analysis to static boundaries, where the potential vanishes inside the domain and is infinite elsewhere, with no time-dependent components. The simplest instance is the circular billiard, described in polar coordinates by $R_b(\theta)=1$, with fully integrable dynamics. The specular reflection of a particle inside this domain is captured by the mapping $\theta_{n+1}=\theta_n + \pi - 2\alpha_n$ and $\alpha_{n+1}=\alpha_n$, where $\theta_n$ and $\alpha_n$ represent, respectively, the angular position relative to the origin and the angle relative to the tangent at the $n$-th collision \cite{chernov}. The phase space of the circular billiard, along with a sketch of the boundary and a single orbit after 50 iterations, is presented in Fig.~\ref{Fig1}(a).

This integrability remains when the circular boundary is deformed into an ellipse. The elliptical billiard is described by $R_b(\theta)=(1-e^2)/(1+e\cos(\theta))$, recovering the circle when the eccentricity $e \in [0,1)$ is equal to zero. As straightforward mapping is no longer easily obtained, as a transcendental equation must be solved for $\theta_{n+1}$. The phase space for $e=0.3$ is shown in Fig.~\ref{Fig1}(b), with islands of stability along the previously identified invariant curves. These curves correspond to rotating orbits (blue trajectory in panel b), while the islands correspond to librating orbits (red trajectory in the same figure). The division between these regimes is represented by a separatrix curve, which turns into an stochastic layer when time-dependence is introduced \cite{berry}. Despite the emergence of distinct phase-space structures, the elliptical billiard remains integrable for all values of $e$ \cite{onthe}.

 On the other hand, a similar deformation can lead to markedly different dynamics. The oval billiard, given by $R_b(\theta)=1+\varepsilon \cos(p\theta)$, also recovers the circle when the deformation $\varepsilon \in [0,1)$ equals zero. However, unlike the ellipse, is not integrable for any $\varepsilon \neq 0$ \cite{onthe}. The parameter \textit{p} introduces an additional deformation, leading to the "oval-like" billiard for $p\neq 1$. This value must also be integer to avoid discontinuities in the boundary.  The phase space, presented in Fig.~\ref{Fig1}(c), exhibits the aforementioned regular structures alongside a chaotic sea bounded by the invariant curves. As the control parameter is increased, the chaotic sea expands, eventually leading to a fully chaotic phase space once convex pieces appear on the boundary. This transition is marked by a critical value $\varepsilon_c=1/(1+p^2)$ corresponding to a local change in curvature from positive to negative \cite{critic}.

 Despite all these considerations being well established for each individual case, their interplay presents several open problems. The elliptical-oval billiard, given by \cite{onthe}
\begin{equation}
R(\theta,p,q,e,\varepsilon)=\frac{1-e^2}{1+e\cos(q\theta)}+\varepsilon \cos (p\theta).
\label{front}
\end{equation}
fills this role introducing an elliptical component onto the mixed dynamics of the oval billiard. As mentioned, $\varepsilon$ and $e$ control the boundary deformations, recovering the circular billiard when $\varepsilon=e=0$. The parameter $q$, similarly to $p$, must also be an integer and introduces additional modulations to the ellipse. For $q \neq 1$, the "elliptical-like" billiard is no longer fully integrable, exhibiting a mixed phase space akin to the oval case \cite{livrodenisspringer}.

The resulting system is more than a simple overlay of the elliptical and oval cases: the interference between the different amplitude modulations of the two terms in Eq.~(\ref{front}) can lead to new dynamics emerging in phase space, as well as expand the range of boundaries and systems that can be modeled. Consequently, the previously established analytical expression for $\varepsilon_c$ is no longer valid for this combined system, leaving open questions regarding the transition to global chaos.

We investigate the change in curvature to derive a new analytical expression for the critical parameter $\varepsilon_c$ in this system. The result is validated numerically by searching for the critical value at which no regular curves remain. This is accomplished using Slater's theorem, which states that for a quasiperiodic orbit, there are at most three recurrence times, with the largest equal to the sum of the other two \cite{slater}. This approach was previously successfully applied to the oval billiard \cite{joelson}, providing a solid numerical confirmation of the expression. As expected, the introduction of additional deformations and the increase of the elliptical component lower the critical threshold for global chaos. We also investigate a regime where the two deformation components are in phase, revealing that the elliptic contribution progressively suppresses chaos, leading to the restoration of invariant curves and periodic orbits. These results allow us to better understand how boundary deformations shape phase-space structures in billiards.

This paper is organized as follows: Section \ref{Sec2} describes the model and the mapping for the static elliptical-oval billiard. In Section \ref{Sec3} we discuss the effects of introducing the elliptical component into the oval billiard phase space. The analytical expression for $\varepsilon_c$ in this new scenario is obtained, showing the decrease of this value as $e$, $p$, and $q$ increase. Section \ref{Sec4} details the method based on Slater's theorem, associating it with the determination of the existence and location of invariant curves, and applying it to the numerical validation of the results presented in Section \ref{Sec3}. Section \ref{Sec5} is dedicated to the specific case $p = q$, where the two deformations are in phase and the elliptical contribution reverses its role, restoring regular structures in phase space. We propose a first-order approximation for this scenario and validate it using the same numerical method. Section \ref{Sec6} presents the final remarks, summary, and conclusions

\section{\label{Sec2} Model and mapping}

\begin{figure*}[htb]
\hspace{-20pt}
\includegraphics[width=1.44\columnwidth]{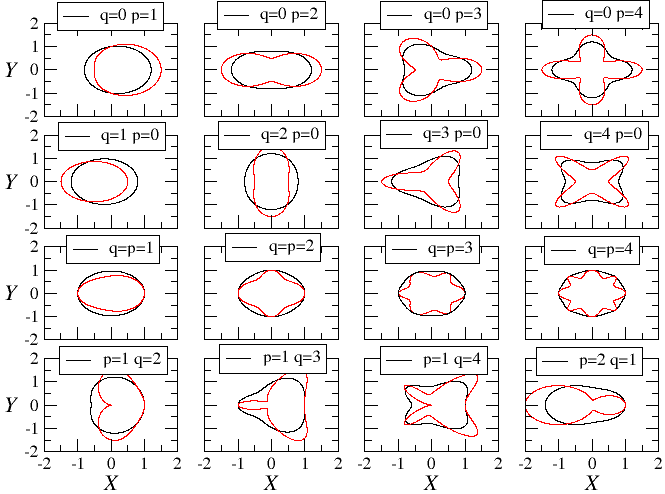}% Here is how to import EPS art
\caption{\label{Fig2} Geometry of the boundary for the elliptical-oval billiard for different combinations of $p$ and $q$, $\varepsilon=e=0.2$ (black) and $\varepsilon=e=0.5$ (red). }
\end{figure*}

In this section we detail the model and mapping for the static elliptical-oval billiard. Inside the region bounded by the curve given in  Eq.~(\ref{front}), a particle or, set of non-interacting ones, undergoes elastic and specular collisions with the boundary. Fig.~\ref{Fig2} displays 32 sets of boundaries obtained from different combinations of the control parameters. 

The integer values of $p$ and $q$ control the number of lobes observed on the boundary. When both values are nonzero, the result is not simply the direct sum of the number of nodes, but rather a consequence of the interference between these different folds. This effect is particularly noticed at the $p\neq q$ boundaries shown in Fig.~\ref{Fig2}.  Simultaneously, $\varepsilon$ and $e$ control the amplitude of such deformations associated with the oval and the ellipse, respectively. In Fig.~\ref{Fig2} this can be observed by comparing the black ($\varepsilon=e=0.2$) and red ($\varepsilon=e=0.5$) boundaries for fixed $p$ and $q$. On the other hand, while the boundary behaves as simple sum in the number of nodes for $p=q\neq0$ and $\varepsilon=e=0.5$, in the low-deformation regime the two deformations appear to nearly cancel each other through interference.

\begin{figure}[htb]
\includegraphics[width=0.58\columnwidth]{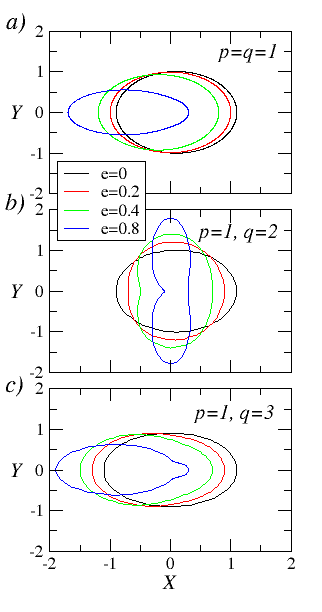}% Here is how to import EPS art
\caption{\label{Fig3} Geometry of the boundary for the elliptical-oval billiard for $\varepsilon=0.1$ with a)$p=q=1$, b) $p=1,q=2$ and c) $p=1,q=3$. Different colors indicate progressively higher values of $e \in [0,0.8]$ }
\end{figure}

\begin{figure}[htb]
\includegraphics[width=\columnwidth]{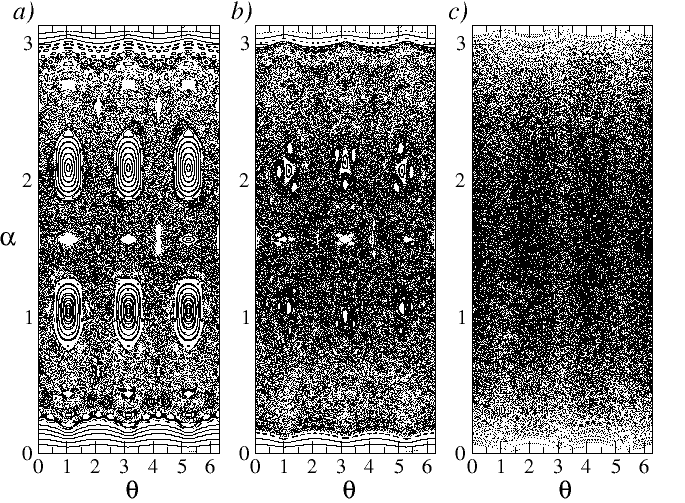}% Here is how to import EPS art
\caption{\label{Fig4} Phase spaces for the oval and elliptical-oval billiards for different sets of control parameters with $p=3$. Panels $a)$ and $c)$ correspond to the strictly oval case ($q=e=0$), with $\varepsilon=0.05<\varepsilon_c$ and $\varepsilon=0.15>\varepsilon_c$, respectively. Panel $b$ shows the elliptical-oval billiard with $q=1$ and $e=\varepsilon=0.05$, illustrating the expansion of the chaotic sea observed in panel $a)$ when the elliptical component is introduced.}
\end{figure}

Alternatively, we can visualize Eq.~(\ref{front}) as an oval billiard with the elliptical component acting as a perturbation. This approach, illustrated in Fig.~\ref{Fig3}, allows us to notice how, even in a oval system below the critical value $\varepsilon_c$, a small amplitude $e$ is sufficient to break the symmetry of the boundary and,  consequently, enhance the chaotic dynamics of the system.
To better elucidate the changes in particle dynamics suggested by Figs.~\ref{Fig2} and~\ref{Fig3}, we now turn our attention to the resulting phase space. The particle position at the collision $n$ can be written as
\begin{gather}
    X(\theta_n)=R(\theta_n)\cos(\theta_n), \\
    Y(\theta_n)=R(\theta_n)\sin(\theta_n).
\end{gather}
As mentioned, $\theta_n$ indicates the angular position of the particle measured from the origin. The angle between the tangent and the boundary at this position is
\begin{equation}
    \phi_n=\arctan\left[\frac{Y'(\theta_n)}{X'(\theta_n)}\right],
    \label{fi}
\end{equation}
and the expressions for $X'(\theta_n)$ and $Y'(\theta_n)$ are written as:
\begin{gather}
    X'(\theta_n)=\frac{dR(\theta_n)}{d\theta_n}\cos(\theta_n) - R(\theta_n)\sin(\theta_n), \\
    Y'(\theta_n)=\frac{dR(\theta_n)}{d\theta_n}\sin(\theta_n) + R(\theta_n)\cos(\theta_n).
    \label{1deriv}
\end{gather}
As the velocity remains constant between successive collisions, the trajectory is
\begin{equation}
    Y(\theta_{n+1})-Y(\theta_n)=\tan(\alpha_n+\phi_n)[X(\theta_{n+1})-X(\theta_n)].
\end{equation}
The variable $\alpha_n$ denotes the angle between the trajectory and the tangent vector to the boundary at $\theta_n$. The two-dimensional nonlinear mapping for the pair $(\theta_n,\alpha_n)$ is then \cite{onthe}
\begin{equation}
\begin{aligned}
h(\theta_{n+1}) &= R(\theta_{n+1})\sin(\theta_{n+1})-Y(\theta_n)
-\tan(\alpha_n+\phi_n) \\
&\qquad \times [R(\theta_{n+1})\cos(\theta_{n+1})-X(\theta_n)] \\
\alpha_{n+1} &= \phi_{n+1}-(\alpha_n+\phi_n)
\end{aligned}
\label{map}
\end{equation}
Unlike in the circular case, $\theta_{n+1}$ must be obtained numerically from the solution $h(\theta_{n+1})=0$. Fig.~\ref{Fig4} presents a comparison between the phase space generated Eq.~(\ref{map}) for a strictly oval billiard and for the elliptical-oval for the same values of $p$ and $\varepsilon$.

 Fig.~\ref{Fig4} \textit{a)} shows the mixed phase space of the oval billiard ($p=3$, $\varepsilon=0.05$), with chaotic regions surrounded by regular structures such as periodic islands and invariant spanning curves. Fig.~\ref{Fig4} \textit{b)} shows the expected increase in this chaotic regime when a elliptical component is introduced under the same parameters ($p=3$, $\varepsilon=e=0.05$, $q=1$). In this case, part of the regular structures are destroyed, but the mixed phase space behavior still prevails, similarly to what would be observed through an small increase in $\varepsilon$ in the oval case. 

 In  Fig.~\ref{Fig4} \textit{c)} all of the regular structures present on the phase space are destroyed. Once the presence of regular orbits is favored by the existence of concave pieces on the boundary,  convex pieces have the opposite effect. Once present, certain regions of the billiard become increasingly difficult to access. This leads to the destruction of the invariant spanning curves observed at the top and bottom of the phase space.  Once the last invariant curve is destroyed the phase space becomes fully chaotic.

The exact value of $\varepsilon$ at which this occurs, denoted $\varepsilon_c$, can be predicted by the expression \cite{critic} 
\begin{equation}
    \varepsilon_c=\frac{1}{(1+p^2)} \ \ \ \text{$p>1$},
    \label{criticovoid}
\end{equation}
whose derivation is detailed in Section \ref{Sec3}.  As mentioned, for $p=3$,  Fig.~\ref{Fig4} \textit{c)} illustrates the results for $\varepsilon=0.15>\varepsilon_c$. The increase of the chaotic sea after the elliptical component is introduced, together with the absence of an analytical expression such as Eq.~(\ref{criticovoid}) for the elliptical-oval billiard, constitutes the main motivation of this work. In Section \ref{Sec3}, the solution for the elliptical-oval billiard is presented, following the deduction of Eq.(9) and confirming the expected influence of the new parameters $q$ and $e$ on the value of $\varepsilon_c$.

\section{\label{Sec3} Effects of the elliptical component on the critical value $\varepsilon_c$}
\begin{figure}
    \centering
    \includegraphics[width=\linewidth]{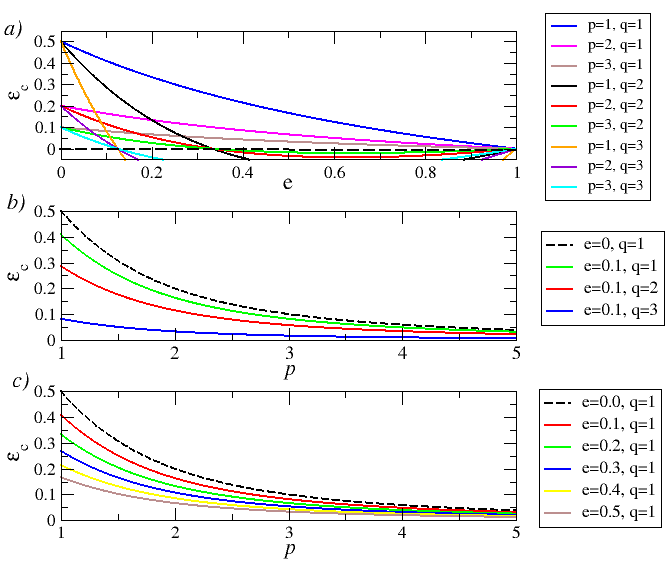}
    \caption{Critical parameter $\varepsilon_c$ as a function of (a) $e$ for different $p$ and $q$, (b) $p$ for different $q$ with fixed $e$, and (c) $p$ for different $e$ with fixed $q$. In panels b) and c), dashed black curves represent the strictly oval case.}
    \label{fig5}
\end{figure}

Returning to the analysis of the parameter $\varepsilon$ and its role in the destruction of periodic structures, an increase in $\varepsilon$ leads to the emergence of chaotic regions surrounding the islands of periodicity, or KAM islands. It is possible to obtain, from the boundary characteristics, the critical value of $\varepsilon$ that leads to the destruction of the last regular structure, after which the phase space is completely overtaken by a chaotic sea. This result is obtained by analyzing the curvature
\begin{equation}
    \kappa(\theta)=\frac{X'(\theta)Y''(\theta) - X''(\theta)Y'(\theta)}{[X'^2(\theta) + Y'^2(\theta)]^{\frac{3}{2}}},
    \label{kapa}
\end{equation}
which marks the transition from positive and strictly convex ($\kappa>0$) to negative ($\kappa<0$), associated with non-convex pieces \cite{critic}. The first derivatives are given by Eq.(5) and (\ref{1deriv}), while $X''(\theta)$ and $Y''(\theta)$ are
\begin{gather}
    X''(\theta)=\frac{d^2R(\theta)}{d\theta^2}\cos(\theta) - 2\frac{dR(\theta)}{d\theta}\sin(\theta)-R(\theta)\cos(\theta),\\
     Y''(\theta)=\frac{d^2R(\theta)}{d\theta^2}\sin(\theta) - 2\frac{dR(\theta)}{d\theta}\cos(\theta)-R(\theta)\sin(\theta).
\end{gather}
Replacing the expression for the oval boundary and solving $\kappa=0$ leads to $\varepsilon_c=1/(1+p^2)$. Applying the same procedure to Eq.~(1) yields the numerator of $\kappa$:
\begin{gather}
   \kappa_{num}= \frac{1}{(e+1)^4}\Bigg[(e+1)^2 \left(-e^2+(e+1) \varepsilon +1\right)^2+\nonumber\\\left(-e^2+(e+1) \varepsilon +1\right) \Big((e+1)^3 \varepsilon  p^2\nonumber\\-\frac{1}{2} (-2 e-2) e \left(e^2-1\right) q^2\Big)\Bigg].
\end{gather}
Solving $\kappa_{\text{num}}=0$ for $\varepsilon$ leads to
\begin{equation}
     \varepsilon_c=\frac{(e-1) \left(e \left(q^2-1\right)-1\right)}{(e+1) \left(p^2+1\right)}.
     \label{eqcrit}
\end{equation}
This solution not only recovers the strictly oval case when $e=0$ but also shows that the elliptical parameters influence the destruction of periodic structures, despite the elliptical billiard itself being integrable (when $q=1$). Fig.~\ref{fig5} shows the behavior of Eq.~(\ref{eqcrit}) for different scenarios: $\varepsilon_c$ vs. $e$ for different combinations of $p$ and $q$ (panel a); $\varepsilon_c$ vs. $p$ for different values of $q$ with fixed $e$ (panel b); and $\varepsilon_c$ vs. $p$ for fixed $q$ and different values of $e$ (panel c). The comparison between the curves in panels a) and b) with the results for the strictly oval case (dashed black curves) confirms that $q$ and $e$, once introduced, accelerate the rate at which the system approaches the critical scenario $\varepsilon > \varepsilon_c$, with complete destruction of all regular structures. These same panels also confirm that the influence of $q$ is stronger than that of $e$ in this process, which is expected from the effects of each parameter on the boundary shape, as discussed in Section II. 

The results in Fig.~\ref{fig5}(a) also show expected values of $\varepsilon_c$ below zero, which is not physically possible. We therefore restrict ourselves to positive values of $\varepsilon$ and find the exact value of $e$ where $\varepsilon_c = 0$:
\begin{equation}
e_c = \frac{1}{(q^2 - 1)}, \qquad q > 1.
\label{e}
\end{equation}
Equation (\ref{e}) indicates the configuration at which, regardless of the value of $\varepsilon$, all regular structures are already destroyed, simultaneously providing the critical value $e_c$ for the strictly elliptical-like case (with $q > 1$), since the elliptical case ($q = 1$) remains always fully integrable.

\begin{figure}[H]
    \centering
    \includegraphics[width=\linewidth]{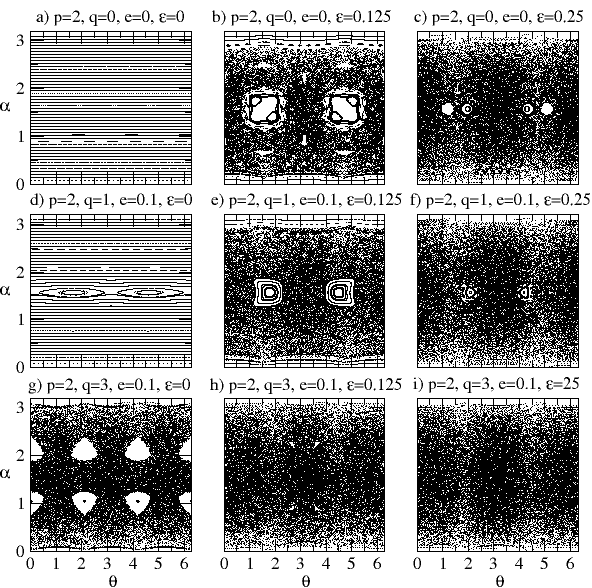}
    \caption{Phase spaces for $p=2$ and different values of $\varepsilon$, $q$ and $e$.}
    \label{fig6}
\end{figure}

Lastly, a set of different phase space portraits is presented in Fig.~\ref{fig6}. The first row shows the known results for the oval case ($q = e = 0$): an initially integrable phase space that becomes mixed once $\varepsilon \neq 0$, with the last invariant curve being destroyed for $\varepsilon > \varepsilon_c = 0.2$. The second row presents the results after an elliptical component ($e = 0.1$, $q = 1$) is introduced. For the same values of $\varepsilon$, new structures emerge and, by directly comparing panels (b) and (e), we see an increase of the chaotic sea, as well as the destruction of invariant curves previously present close to $\theta = \pi$. The last row shows the even greater impact of an elliptical component with $q = 3$. Even without an oval component ($\varepsilon = 0$), the phase space already presents a significant chaotic sea, consistent with a value of $e$ close to $e_c = 0.125$. For values of $\varepsilon$ as low as $0.125$ (panel h), no more regular structures are identified. These phase space portraits are consistent with the expected behavior described by Eqs.~(14) and~(15).

 We now turn to a numerical method based on Slater's theorem in order to confirm the exact value at which the last invariant curve is destroyed, further validating the expressions found. 
\section{\label{Sec4} Numerical results}

The three-gap theorem, also known as the Steinhaus conjecture, is a classical geometric result stating that if points are placed on a circle at angles $\theta, 2\theta, \dots, N\theta$ with $\theta$ irrational, the distances between adjacent points on the circle can assume at most three distinct values. Although foundational work was conducted by Slater as early as 1950 \cite{Slater1950}, the theorem was first formally proven in 1958 by S{\'o}s \cite{Sos1958}, Sur{\'a}nyi \cite{Suranyi1958}, and {\'S}wierczkowski \cite{Swierczkowski1959}. Later, Slater showed that an equivalent property holds for irrational rotations when considering recurrence to an interval \cite{slater}. The same result holds for the dynamics of invariant curves, since their quasi-periodic motion is related to a simple rotation on the circle \cite{slaterref}. Therefore, if we mark a small interval $\delta$ along an orbit and track when the trajectory revisits it, the time intervals between successive visits can assume at most three values satisfying $\Gamma_3 = \Gamma_1 + \Gamma_2$. This behavior serves as a signature of motion on an invariant curve.Therefore, if we mark a small interval $\delta$ along an orbit and track when the trajectory revisits it, the time intervals between successive visits can assume at most three values satisfying $\Gamma_3 = \Gamma_1 + \Gamma_2$. This behavior serves as a signature of motion on an invariant curve.

The method based on this result consists of calculating the recurrence times for a given orbit in terms of the coordinates $(\theta, \alpha)$\cite{joelson}. The recurrence times, in this context, correspond to the number of iterations necessary for an orbit to return to an interval $\delta$ close to its starting point. If only three recurrence times are found and they obey Slater's theorem, we conclude that the point belongs to an invariant curve. The procedure is repeated for a fixed value of $\theta$, scanning different initial values of $\alpha$ in steps of $\Delta\alpha$ across the phase space.

The application of this method enables a more precise identification of invariant curves, including those that might not be visible in the initial phase space due to insufficient initial conditions or iterations. Fig. 7 shows, in red, the coordinates that satisfy Slater's theorem, which correspond to invariant curves for $\delta = 10^{-5}$. The last two points in each figure represent the last two invariant curves found, which were not identified in the initial phase space. Table 1 provides the coordinates of each of these curves, along with their corresponding recurrence times, with the largest one equal to the sum of the other two.

\begin{figure}[h]
    \centering
    \includegraphics[width=1.1\linewidth]{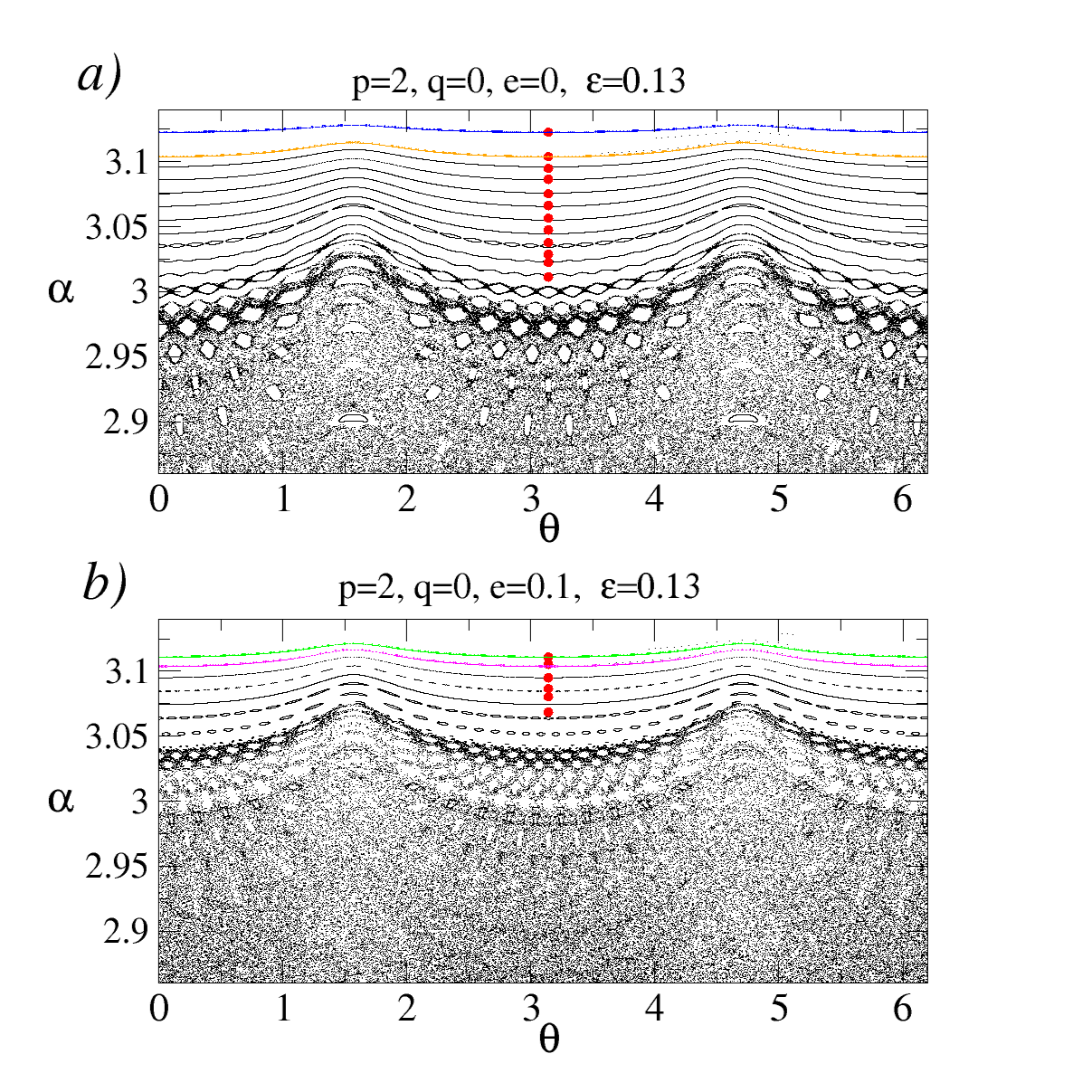}
    \caption{Phase spaces for $p=2$ and different values of $\varepsilon$, $q$ and $e$, The last two invariant curves found are colored in blue, yellow, magenta and green. The four of them were not identified in the initial phase space}
\end{figure}

\begin{table}[h]
\caption{\label{tab:table1}Recurrence times for $\delta=10^{-5}$ and coordinates for the colored invariant curves in Fig. 7.  }
\begin{ruledtabular}
\begin{tabular}{l c c c c}
Curve & $\alpha$\tablenote{$\theta$ is fixed in $\pi$}& $\Gamma_1$ & $\Gamma_2$ & $\Gamma_3$ \\
\hline
Blue (panel a) & 3.1227 & 13173 & 7579 & 20752 \\
Orange (panel a) & 3.1038 & 9203 & 12000 & 21203  \\
Green (panel b) & 3.1110 & 9972 & 17479 & 27451 \\
Magenta (panel b) & 3.1049 & 16900 & 21942 & 38842 \\
\end{tabular}
\end{ruledtabular}
\end{table}

Fig. 7 also allows for a visualization of the destruction of invariant curves provoked by the elliptical component $e$. Comparing panels (a) and (b), for small intervals of two phase spaces with $p=2$ and different values of $\varepsilon$, a significant decrease in the number of curves found is observed, with the remaining ones seemingly compressed into a smaller region.

We now extend this approach to determine the critical parameter $\varepsilon_c$ after which all invariant curves are destroyed. The method consists of gradually increasing the control parameter and searching for invariant curves. Once no point satisfying the criteria is found, indicating that all invariant curves have been destroyed, we select a new value of $\varepsilon$ between the last value where regular structures were identified and the first where they were absent, halving the interval. This process is repeated iteratively. The largest value of $\varepsilon$ for which an invariant curve is still detected is then identified as the critical parameter.

The proposed method is implemented for the elliptical-oval billiard, with the results presented in Fig. 8 for $\delta = 10^{-6}$. The symbols represent the numerical values, while the curves represent the expected behavior from Eq. (14). These results show remarkable agreement between theoretical and numerical predictions for different combinations of $p$ and $q$. Both the theoretical and numerical results also confirm the effect of the elliptical component on $\varepsilon_c$, at rates that depend on $p$ and $q$. Furthermore, the dotted black line allows for visualization of the behavior predicted by Eq. (15). In this regime, the regular structures were all destroyed even in the absence of an oval deformation ($\varepsilon = 0$). The absence of numerical values found below the line further validates Eq. (15). Thus, the method provides insight into the transition to chaos in this combined system, previously unexplored, extending the results obtained for the strictly oval case \cite{joelson}.

\begin{figure}[h]
    \centering
    \includegraphics[width=\linewidth]{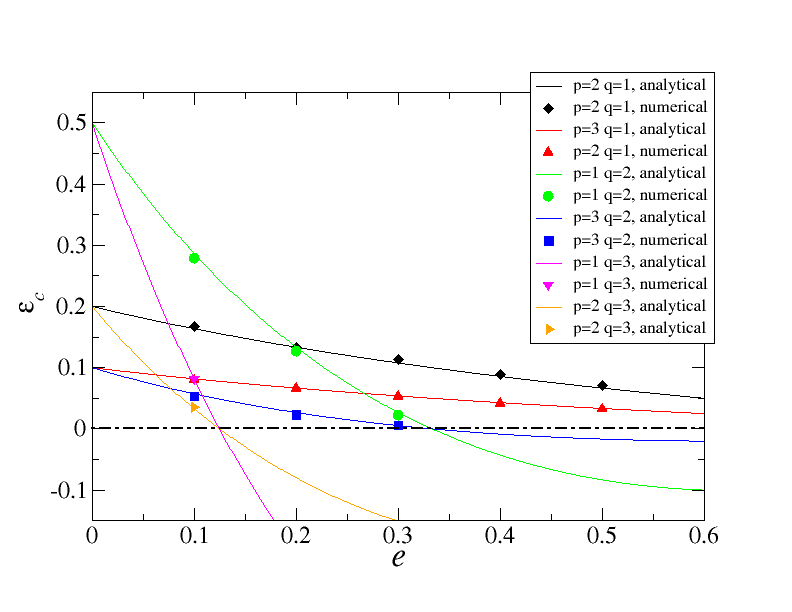}
    \caption{Critical parameter $\varepsilon_c$ as a function of $e$ for different values of $p$ and $q$, with $\delta = 10^{-6}$. The symbols represent the numerical results and the curves are obtained from Eq. (14).The dotted black line represents the regime described by Eq. (15), where $\varepsilon_c = 0$}
\end{figure}
Despite the general agreement between the numerical and theoretical results in the previous cases, the scenario where $p = q$ deserves special attention. As shown in Fig. 9, invariant curves are found in the phase space for $\varepsilon = 0.55$. This value is much larger than that predicted by Eq. (14) for $p = q = 1$ and $e = 0.1$ ($\varepsilon_c = 0.4\overline{09}$). Three invariant curves were found to obey Slater's theorem, with their recurrence times presented in Table 2. 

\begin{figure}[h]
    \centering
    \includegraphics[width=\linewidth]{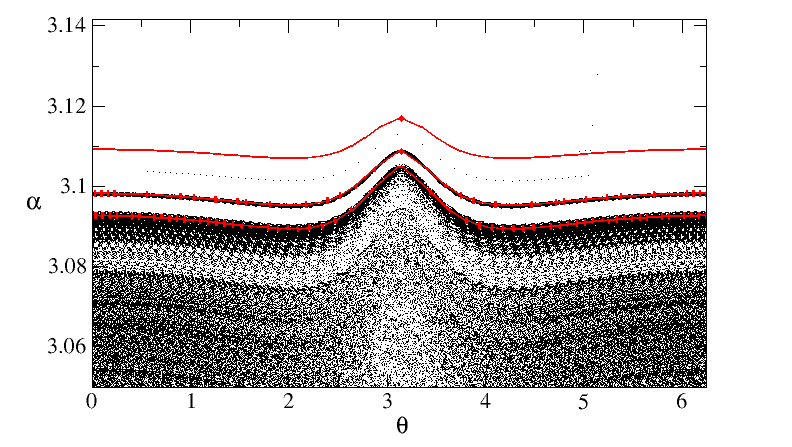}
    \caption{Phase space for $p=q=1$ $e=0.1$ and $\varepsilon=0.55$. Three invariant curves where found to obey Slater's theorem, with the recurrence times presented in Table II. The expected critical value given by Eq.(14) for this set of control parameters is $\varepsilon_c=0.4\overline{09}$.}
\end{figure}

\begin{table}[h]
\caption{\label{tab:table1}Recurrence times for $\delta=10^{-5}$ and coordinates for the red invariant curves in Fig. 9.  }
\begin{ruledtabular}
\begin{tabular}{l c c c}
$\alpha$\tablenote{$\theta$ is fixed in $\pi$}& $\Gamma_1$ & $\Gamma_2$ & $\Gamma_3$ \\
\hline
3.1046 & 9009 & 25137 & 34146 \\
3.1087 & 16969 & 5822 & 22791  \\
3.1169 & 2470 & 5415 & 7885 \\
\end{tabular}
\end{ruledtabular}
\end{table}

This behavior occurs due to the interference between the elliptic and oval components, which are in phase when $p = q$. In this context, the introduction of the elliptic component acts to increase, rather than decrease, the value of $\varepsilon_c$. In Section V we will discuss this scenario further, proposing an alternative solution for $p = q$.

\section{\label{Sec5} Solution for the $p=q$ case}
An initial visualization of the results for $p = q$ was previously shown in Fig. 2. In the third row, for $p = q = 1, 2, 3,$ and $4$, one can observe that for small deformations ($\varepsilon = e = 0.2$), the resulting boundary shapes approximate a circle. Increasing these values to $\varepsilon = e = 0.5$ leads to the expected $p+q$ lobes. This indicates that, even in this scenario, a critical value $\varepsilon_c$ exists at which convex pieces appear on the boundary. However, the results in Fig. 9 and Table 2 show that Eq. (14) fails to predict this critical value for the specific case $p = q$.

A direct comparison between $p = 1, q = 2$ and $p = q = 1$ is presented in Fig. 10. The indigo phase space, which is completely chaotic for $\varepsilon = 0.5$ and $e = 0.3$, is expected as $\varepsilon_c \approx 0.196$ for these parameters. This chaotic sea is also a consequence of the heart-shaped boundary found for $p = 1, q = 2$, shown above the phase space, with a clear convex piece. In contrast, the same values of $\varepsilon$ and $e$ lead to a very different phase space configuration for $p = q = 1$ (black). Here, the chaotic sea coexists with periodic islands and invariant curves, even though $\varepsilon$ remains above the critical value for this parameter set ($\varepsilon_c \approx 0.269$). The boundary now appears unfolded, taking on an oval shape even for significant values of $\varepsilon$ and $e$, as shown above the black phase space.

\begin{figure}[h]
    \centering
    \includegraphics[width=\linewidth]{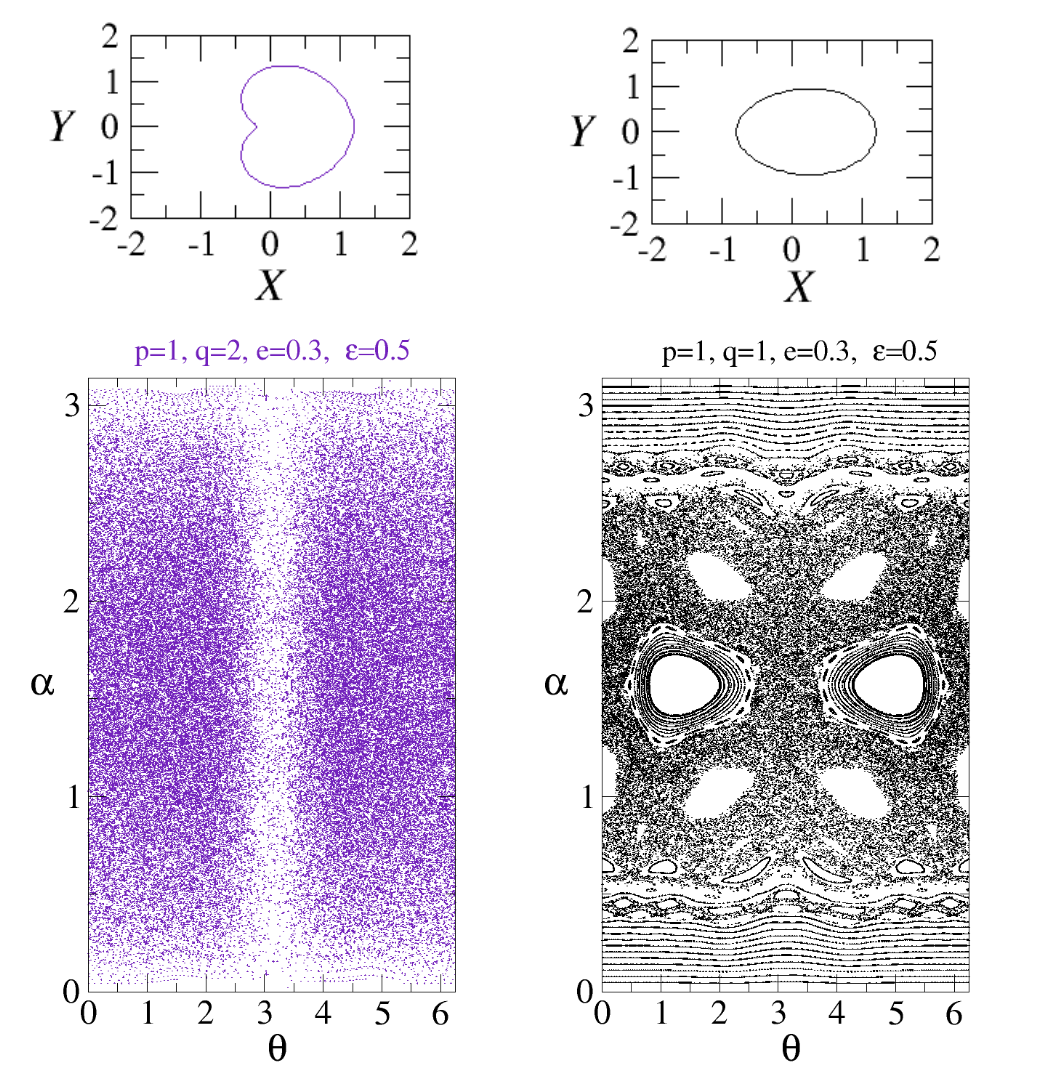}
    \caption{Boundaries and respective phase spaces for $e=0.3$, $\varepsilon=0.5$. The results for $p=1,q=2$ and $p=q=1$ are presented in indigo and black, respectively.}
    \label{fig10}
\end{figure}

This behavior can be explained by considering the superposition of the oval and elliptical components on the boundary. In Fig. 11, the example for $p = q = 3$ is presented. In panel a), the boundaries for the strictly oval (red) and strictly elliptical (black) cases are mirrored images of each other. Their overlap in each section results in the blue boundary, which exhibits deformations of smaller amplitude. This effect is evident in the phase spaces, colored according to their corresponding boundary. Panels b) and c) show completely chaotic phase spaces, with $\varepsilon > \varepsilon_c$. Panel (d), in contrast, presents a mixed phase space, with newly appearing invariant curves and periodic islands throughout. This result allows us to interpret the introduction of the elliptical (or oval) component as a mechanism for suppressing chaos in the phase space. Therefore, the behavior is the inverse of that observed for the case $p \neq q$, and a new analytical solution is required.

\begin{figure}[h]
  \centering
    \centerline{\includegraphics[width=1\linewidth]{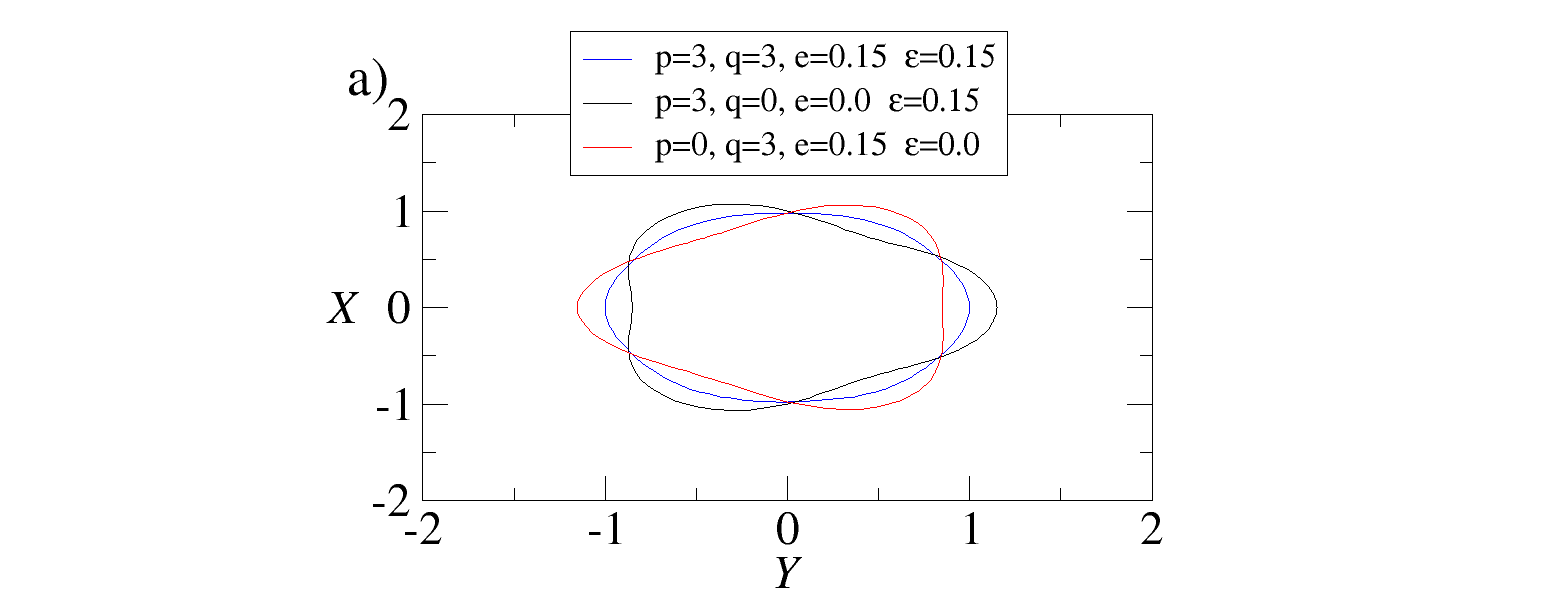}}
    \centerline{\includegraphics[width=1\linewidth]{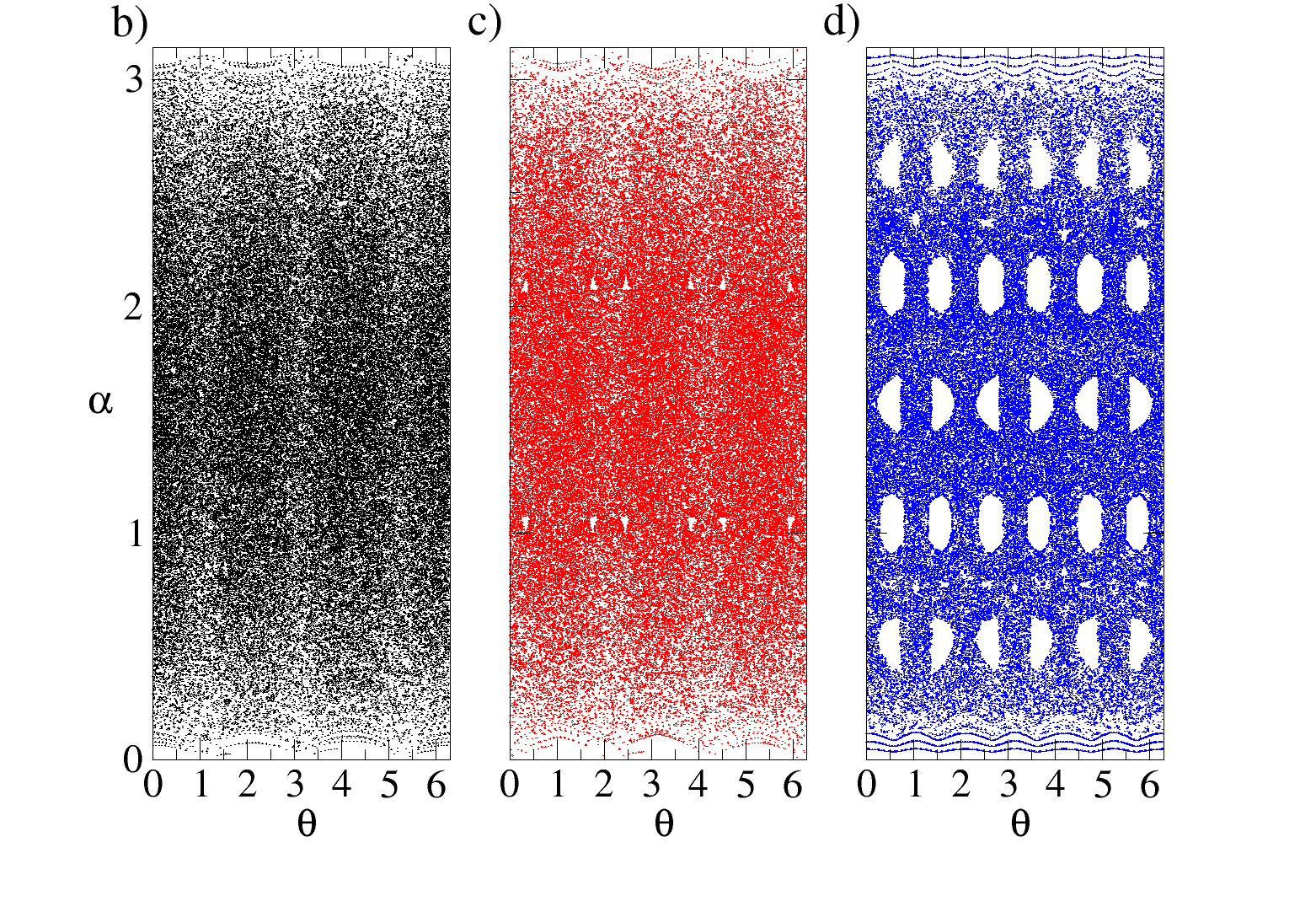}}    
  \caption{\label{fig11} a) Boundaries for the strictly oval case (red), strictly elliptical case (black), and their superposition (blue), showing the partial cancellation of deformations. The associated phase spaces are presented in panels b)-d) and colored accordingly.}
\end{figure}

We rewrite the elliptical component of Eq. (1) using the expansion 
\begin{equation}
    \sum^{\infty}_{k=0} \left(-e\cos(q\theta)\right)^k = 1 - e\cos(q\theta) + e^2\cos^2(q\theta) + \cdots
\end{equation}
Thus, for $p = q$, $R(\theta)$ becomes:
\begin{gather}
R(\theta) = 1 - e\cos(q\theta) + \mathcal{O}(e^2) + \varepsilon \cos(p\theta) \nonumber\\= 1 + (\varepsilon - e)\cos(p\theta) + \mathcal{O}(e^2),
\end{gather}
where $\mathcal{O}(e^2)$ represents higher-order contributions. A straightforward comparison with the strictly oval case leads to $\varepsilon - e + \mathcal{O}(e^2) = \frac{1}{1+p^2}$, or simply
\begin{equation}
\varepsilon_c = e + \frac{1}{1+p^2}
\end{equation}
for the first order approximation.

The critical value $\varepsilon_c$ is once again obtained through the same method based on Slater's theorem, presented in Section IV. Fig. 12 shows a comparison between these numerical results, Eq. (14), and Eq. (18). As expected, Eq. (14) does not accurately predict the critical value for $p = q$. Eq. (18), on the other hand, shows better agreement with the numerical results for $p = q = 1, 2,$ and $3$. The first-order approximation is therefore validated, and the complete solution for the elliptical-oval case is given by

\begin{equation}
    \varepsilon_c = 
\begin{cases}
\dfrac{(e-1)\left[e\left(q^2-1\right)-1\right]}{(e+1)\left(p^2+1\right)}, & \text{if } p \neq q \\[10pt]
e + \dfrac{1}{1+p^2}, & \text{if } p = q
\end{cases}
\end{equation}

\begin{figure}[H]
\centering
\includegraphics[width=\linewidth]{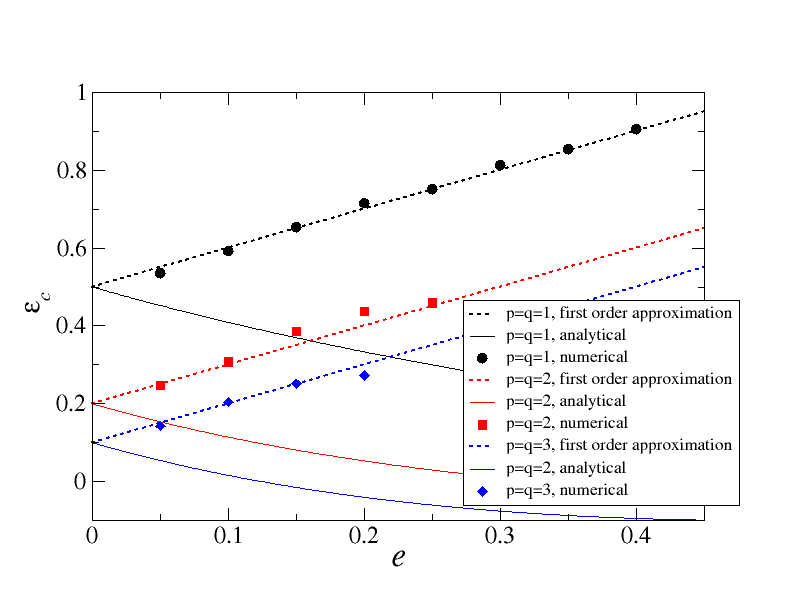}
\caption{\label{fig12} Comparison between numerical and theoretical values of $\varepsilon_c$ for the $p = q$ case. Symbols represent numerical results obtained via Slater's theorem, the solid curves correspond to Eq. (14), and the dashed curves represent the first-order approximation given by Eq. (18) for $p = q = 1, 2, 3$. }
\end{figure}

\section{\label{Sec6} Summary and conclusions}

We investigated the transition from integrability to chaos in a family of billiards combining elliptical and oval boundary deformations. Unlike the standard oval billiard, where a single critical parameter $\varepsilon_c = 1/(1+p^2)$ governs the destruction of the last invariant curve, the introduction of an integrable elliptical component introduces a second deformation axis, leading to more complex phase-space dynamics. We derived an analytical expression for $\varepsilon_c$ in the elliptical-oval billiard, which recovers the oval case when $e = 0$ and shows how the elliptical parameters $e$ and $q$ accelerate the destruction of regular structures. A solution for the critical value $e_c$ for the elliptical-like boundary was also found, predicting complete destruction of regular structures without the need for the oval component. Numerical validation using Slater's theorem demonstrated excellent agreement with the analytical solution for various combinations of $p$, $q$, and $e$.

A previously unexplored regime was identified when the deformation components are in phase ($p = q$). In this scenario, the elliptical component suppresses chaos rather than enhancing it, leading to the restoration of invariant curves and periodic orbits. A first-order analytical approximation was derived and numerically validated. This behavior arises from interference between the oval and elliptical components, which partially cancel each other's deformations, resulting in a boundary that remains nearly circular even for relatively large amplitudes. Our results demonstrate that the interplay between distinct boundary deformations enriches phase-space organization beyond simple superposition, offering enhanced tunability of chaotic dynamics through multiple control parameters. Future investigations may extend this framework to time-dependent boundaries, dissipative systems, or different boundary shapes, where similar interference effects could give rise to novel dynamical phenomena.

\section*{Acknowledgments and Funding}

We would like to acknowledge financial support from the
Brazilian agencies: Anne Kétri P. da Fonseca received funding from CAPES under Grant Agreement No.~$88887.990665/2024-00$.
Joelson D. V. Hermes thanks Federal Institute of Education, Science and Technology of South of Minas Gerais—IFSULDEMINAS. Edson D. Leonel received funding from Brazilian agencies CNPq under Grant Agreements No.~$301318/2019-0$ and $304398/2023-3$, and from FAPESP under Grant Agreements No.~$2019/14038-6$ and No.~$2021/09519-5$. 
\section*{Author Declarations}
\subsection*{Conflicts of interest}

The authors declare the following financial interests and relationships that could be considered potential competing interests: Edson D. Leonel reports that equipment and supplies were provided by the São Paulo State University, Institute of Geosciences and Exact Sciences. He also reports a relationship with the State of São Paulo Research Foundation involving board membership. Anne Kétri P. da Fonseca and Joelson D. V. Hermes declare no conflicts of interest relevant to the content of this article.
\subsection*{CRediT authorship contribution statement}

\textbf{Anne Kétri P. da Fonseca}: Software, Validation, Formal analysis, Data curation, Writing – Original Draft, Writing – Review \& Editing. 
\textbf{Joelson D. V. Hermes}: Conceptualization, Methodology, Supervision, Formal analysis, Writing – Review \& Editing. 
\textbf{Edson D. Leonel}: Conceptualization, Methodology, Supervision, Writing – Review \& Editing. 

\section*{Data Availability Statement}

The data that support the findings of this study are available from the corresponding author upon reasonable request.
\\

\section*{References}
\bibliography{aipsamp}% Produces the bibliography via BibTeX.

\end{document}